\RequirePackage{silence}
\documentclass[aps,pra,twocolumn,showpacs,amsmath,amssymb,preprintnumbers,superscriptaddress,10pt,longbibliography]{revtex4-2}
\usepackage[bookmarks=false]{hyperref}
\usepackage{graphicx}
\usepackage{SIunits}
\usepackage[capitalise]{cleveref}
	\crefname{section}{Sec.}{Secs.}% APS style uses abbreviations
	\Crefname{section}{Section}{Sections}
\usepackage{color}
\usepackage{tabularray}
\usepackage[italicdiff]{physics}

\definecolor{pink}{RGB}{255,0,255}

\definecolor{red}{rgb}{1,0,0}

\definecolor{green}{RGB}{0,212,17}

\definecolor{orange}{RGB}{255,70,0}

\newcommand{\rev}[1]{\textcolor{black}{#1}}% highlighting revisions for referees is on

\begin{document}

\title{Optical-pumping attack on a quantum key distribution laser source}

\author{Maxim~Fadeev}
\email{mfadeev2022@gmail.com}
\affiliation{Russian Quantum Center, Skolkovo, Moscow 121205, Russia}
\affiliation{ITMO University, St.~Petersburg 197101, Russia}

\author{Anastasiya~Ponosova}
\affiliation{Russian Quantum Center, Skolkovo, Moscow 121205, Russia}
\affiliation{NTI Center for Quantum Communications, National University of Science and Technology MISIS, Moscow 119049, Russia}

\author{Qingquan~Peng}
\affiliation{\mbox{College of Computer Science and Technology, National University of Defense Technology,} \mbox{Changsha 410073, People's Republic of China}}% Country name is spelled this way per APS style

\author{Anqi~Huang}
\affiliation{\mbox{College of Computer Science and Technology, National University of Defense Technology,} \mbox{Changsha 410073, People's Republic of China}}% Country name is spelled this way per APS style

\author{Roman~Shakhovoy}
\affiliation{NTI Center for Quantum Communications, National University of Science and Technology MISIS, Moscow 119049, Russia}
\affiliation{QRate, Moscow 143026, Russia}

\author{Vadim~Makarov}
\affiliation{Russian Quantum Center, Skolkovo, Moscow 121205, Russia}
\affiliation{Vigo Quantum Communication Center, University of Vigo, Vigo E-36310, Spain}
\affiliation{NTI Center for Quantum Communications, National University of Science and Technology MISIS, Moscow 119049, Russia}

%\author[1,2,*]{M.A. Fadeev}
%\author[1,3]{A.A. Ponosova}
%\author[4]{Q. Peng}
%\author[4]{A. Huang}
%\author[3,5]{R. Shakhovoy}
%\author[1,3,6]{V. Makarov}
%
%\affil[1]{Russian Quantum Center, Skolkovo, Moscow 121205, Russia}
%\affil[2]{ITMO University, St.~Petersburg 197101, Russia}
%\affil[3]{NTI Center for Quantum Communications, National University of Science and Technology MISiS, Moscow 119049, Russia}
%\affil[4]{Institute for Quantum Information \& State Key Laboratory of High Performance Computing, College of Computer Science and Technology, National University of Defense Technology, Changsha 410073, People's Republic of China}
%\affil[5]{QRate, Skolkovo, Moscow 143026, Russia}
%\affil[6]{Vigo Quantum Communication Center, University of Vigo, Vigo E-36310, Spain}

%\affil[*]{mfadeev2022@gmail.com}

\date{\today}

\begin{abstract}
We report a new type of vulnerability \rev{based on a physical principle that has not been previously exploited in quantum hacking---optical pumping of a laser} in practical implementations of quantum key distribution (QKD) systems. We show that it is possible to increase the pulse energy of a source laser diode not only by injection-locking it with external light near its emission wavelength of $1550~\nano\meter$, but also by optically pumping it at a much shorter wavelength. We experimentally demonstrate a $10\%$ increase in pulse energy when exposing the laser diode to \rev{continuous-wave (cw) laser light at $1310~\nano\meter$ with a power of $1.6~\milli\watt$} via its fiber pigtail. This \rev{effect} may allow an eavesdropper to steal the secret key. A possible countermeasure is to install broadband optical filters and isolators at the source output and to characterise them during security certification.
%We report a new type of vulnerability \rev{based on new physical principle that has not been used in quantum key distribution (QKD) - optical pumping of laser in practical implementations of QKD systems}. We show that it is possible to increase the pulse energy of a source laser diode not only by injection-locking it by external light near its emission wavelength of 1550~nm, but also by optically pumping it at a much shorter wavelength. We demonstrate 10\% increase in pulse energy when exposing the laser diode to 1310-nm, 1.6-mW cw laser light via its fiber pigtail. This may allow an eavesdropper to steal the secret key. A possible countermeasure is to install broadband optical filters and isolators at the source's output and characterise them during the security certification. 
\end{abstract}

%\setboolean{displaycopyright}{false} % Do not include copyright or licensing information in submission.

\maketitle

\section{Introduction}

Quantum key distribution (QKD) is a technology to generate a true random secret key by remote users using single photons~\cite{bennett1992b,horodecki2008}. The impossibility of compromising QKD protocols via direct measurement of single photons and independence of its security on computation algorithms make QKD an attractive cryptographic tool, especially with the rise of computational technologies. However, attempts at cryptanalysis of practical QKD implementations reveal many imperfections in their hardware implementations, such as bandwidth limitations of modulators~\cite{gnanapandithan2025}, intensity correlation~\cite{trefilov2025}, shot noise measurement~\cite{silva2020} or imperfections in source of quantum states~\cite{tang2016a}. Active quantum-hacking strategies have been proposed that create vulnerabilities in a ``healthy'' system imperceptible for its legitimate users \cite{lo2014,dixon2017,xu2020,sun2022,makarov2024}. To meet hard requirements on cryptographic resistance, practical QKD implementations are studied and countermeasures to ensure physical security of the system hardware are developed and improved.

\medskip% improving page layout

One of the imperfect devices in practical QKD systems is a photon source. Today, strongly attenuated laser pulses from semiconductor laser diodes (LDs) are used instead of true single-photon sources. In several studies, vulnerabilities in QKD are created by seeding Alice's LD by Eve's light injected through the quantum channel \cite{huang2019,pang2020,lovic2023}. The attackers use laser light at about $1550~\nano\meter$, which is near the LD operating wavelength. Injected power in the range of $100$--$160~\nano\watt$ can be sufficient to control the intensity of Alice's pulses~\cite{huang2019,pang2020}. Power as low as $1~\nano\watt$ might be enough to partially control the phase of Alice's pulses \cite{lovic2023}.

In this paper, we investigate the effects of optical pumping~\cite{svelto1998,okamoto2003,guina2017} of the QKD source of coherent radiation by $1310$-$\nano\meter$ illumination from the attacker Eve. This wavelength is a particular case of a broad wavelength band that is absorbed by the InGaAsP crystal within the laser~\cite{levinshtein1999}. This absorption creates an additional inversion population, resulting in an increase in output power. Appendix~\labelcref{sec:supplementary} describes the underlying physical process of the optical-pumping attack and provides a simple theoretical model of it. \rev{Here, we select a wavelength of $1310~\nano\meter$ due to its prevalence and accessibility, which consequently increases the potential risks of the attack.} We demonstrate that QKD source based on a $1550$-$\nano\meter$ gain-switched laser diode is vulnerable to an optical-pumping attack, which results in increase of the energy of pulses emitted by Alice and might compromise the security of the key. 

Our study indicates that the sufficient attenuation of Eve's light entirely mitigates the optical-pumping attack. However, QKD systems may be vulnerable to it due to the spectral selectivity of their passive countermeasures. The analysis of an industrial QKD system~\cite{makarov2024} reveals that active-state-preparation Alice modules may already be immune to this attack. Conversely, passive state-preparation QKD systems without modulators~\cite{zapatero2023, hu2023, lu2023,wang2023} are at greater risk due to the weaker requirements for passive countermeasures. But here we demonstrated the attack in principle at a single wavelength, while an adversary could exploit the entire absorption spectrum of the laser diode material. Effectiveness of the attack at different wavelengths will depend significantly on the practical implementation of a QKD system, including its LD architecture. Thus, comprehensive testing and evaluation are essential to certify QKD systems against this novel attack across the entire LD absorption band.

%We evaluate the vulnerability of an industrial-prototype prepare-and-measure QKD system \cite{makarov2024} to this attack \rev{at $1310~\nano\meter$}.
%\rev{We also note that the optical-pumping attack may be conducted within a whole absorption spectrum of LD semiconductor material. However, its effectiveness at different wavelengths will depend significantly on the practical implementation of a QKD system, including its laser diode (LD) architecture or state preparation. For example systems with passive state preparation~\cite{zapatero2023, hu2023, lu2023} is protected against the "Trojan-horse" attack, it is necessary to do security analysis against this new attack. Thus, it should be verified individually for each real QKD system.}

\begin{figure}
\includegraphics{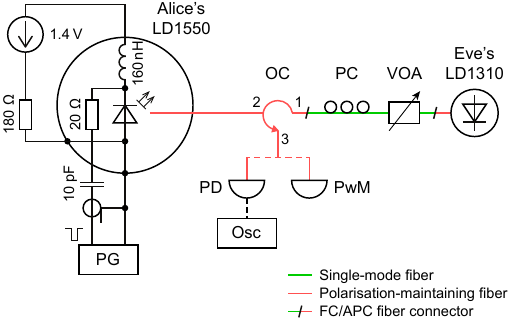}
\caption{Scheme of experiment. LD, laser diode; PG, pulse generator; OC, optical circulator; PD, photodiode; Osc, oscilloscope; PwM, power meter; PC, polarisation controller; VOA, variable optical attenuator. External electrical connections of LD1550 for pulsed operation are shown. Port~3 of OC is interchangeably connected to either PD or PwM.}
\label{fig:experiment}
\end{figure}

\section{Experimental setup}
\label{sec:setup}

Our experimental setup (\cref{fig:experiment}) simulates a hacking scenario when Eve injects light into QKD source at a wavelength significantly shorter than the QKD operating one. A 1550-nm laser diode (LD1550; Agilecom WSLS-934010C4124-82) mimics Alice. However, contrary to the usual industry practice, it lacks a built-in isolator, in order to demonstrate effects using low-power attacker's source (otherwise the isolator would add about $10~\deci\bel$ attenuation at $1310~\nano\meter$). We conduct measurements of QKD source characteristics in both cw and pulsed regimes. In cw regime, LD1550 is powered only by a laboratory power supply (Keysight E3648A). For operating in a gain-switching pulsed mode \cite{svelto2010}, the bias current provided by the power supply is $3~\milli\ampere$, and pulses from the pulse generator (Highland Technology P400) are applied at $10~\mega\hertz$ repetition rate. In this regime, LD1550 produces 700-ps-wide optical pulses, and has an average power of~$14\rev{\pm0.1}~\micro\watt$.

As an attacker, we use a 1310-nm laser diode (LD1310; Nolatech FPL-FBG-1310-14BF) in cw regime. Its emission is injected into LD1550 via a fiber-optic circulator. The output power is controlled with a variable optical attenuator (VOA; OZ Optics BB-100) in a range from $23~\nano\watt$ to $1.6~\milli\watt$ at port~2 of the optical circulator OC. %\rev{Our measurements started from $23~\nano\watt$ and ended at $1.6~\milli\watt$}  
This power is limited by the available maximum power of LD1310. A polarisation controller PC is adjusted to maximise Eve's power at port~2 of OC.

We investigate several characteristics of LD1550 under optical pumping by LD1310: its light--current characteristic and differential quantum efficiency in cw mode, pulse area and shape, and average power in the pulsed regime. The average optical power is measured using an optical power meter (Thorlabs PM400 with a photodiode sensor S154C). The backreflected light at $1310~\nano\meter$ makes a contribution to $1550$-$\nano\meter$ average power measurements at port~3 of OC. We correct for this by measuring the reflected power with LD1550 switched off and subtracting it from the total measured power in each experiment. This correction is stable and is of the same order of magnitude or less than the effects observed. Pulse shape is recorded by a p-i-n photodiode (Laserscom PDI35-10G, $10~\giga\hertz$ bandwidth) connected to an oscilloscope (LeCroy 816Zi, $16~\giga\hertz$ bandwidth). 

First, all the characteristics of LD1550 are measured before exposure. Then, they are characterised under exposure to a constant power level of LD1310 emission. %starting from its maximum available power. 
\rev{In the experiments conducted on the cw 1550-nm LD, we gradually reduce Eve's injected power, beginning from its maximum value. We terminate the experiment when we observe several instances of unchanged characteristics in the 1550-nm LD operation under exposure. Conversely, in the tests of Alice's pulsed source, we gradually increase the 1310-nm power, starting from its minimum value.} 

\section{Experimental results}
\label{sec:results}

We demonstrate how Eve can manipulate the characteristics of Alice's laser by $1310~\nano\meter$ wavelength radiation of different powers. We quantify the influence of Eve's pumping via differential quantum efficiency \cite{cassidy1984, tomiyasu2017}. It indicates how efficiently a laser medium converts injected electron-hole pairs into emitted photons. The theoretical limit for this coefficient is 1. Here we explore how additional optical pumping changes this efficiency.

\begin{figure}
\includegraphics{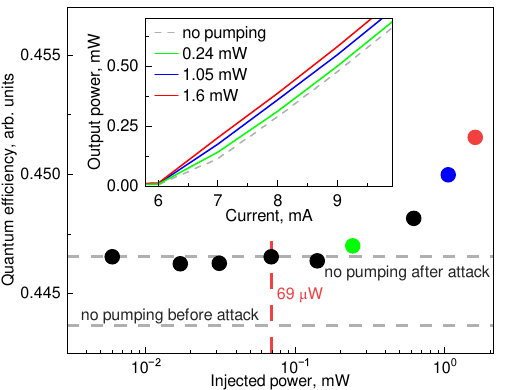}
\caption{Dependence of differential quantum efficiency on the injected cw power of Eve. \rev{``No pumping after attack" shows the level of differential quantum efficiency immediately after exposure.} The inset shows measured light--current characteristics of LD1550 in cw regime.}
\label{fig:quantumeff}
\end{figure}

\Cref{fig:quantumeff} demonstrates differential quantum efficiency of LD1550 depending on the pump power injected through its fiber pigtail. For thus, we measure the output power of LD1550 in cw regime depending on current under different power levels of pumping illumination (inset in~\cref{fig:quantumeff}). We extract the experimental value of optical power -- current slope. We then calculate the differential quantum efficiency
\begin{equation}
\eta = \frac{2e}{\hbar\omega} \dv{P}{I},
\label{eq:quaneff}
\end{equation}
where $e$ is the elementary charge, $\hbar$ is the reduced Planck's constant, $\omega$ is the laser frequency, and $\dv*{P}{I}$ is the power--current slope averaged over $7$ to $25~\milli\ampere$ range.

Our data confirms that, at a fixed bias current, Alice's diode emits higher power under optical pumping. \rev{However, the change in the differential quantum efficiency is less than 1\% in the investigated range of pumping power.} With a decrease in pump power from 1.6 to $0.6~\milli\watt$, the differential quantum efficiency decreases linearly, and then, under the lower pumping power of $140~\micro\watt$ and less, it becomes constant and remains the same even \rev{immediately} after exposure but higher comparing to the pre-exposure level. \rev{It recovers to the initial value within a day.} Both pre- and after-exposure levels are marked with dashed lines in \cref{fig:quantumeff}. Additional research is required for an explanation of this effect.

\begin{figure}
\includegraphics{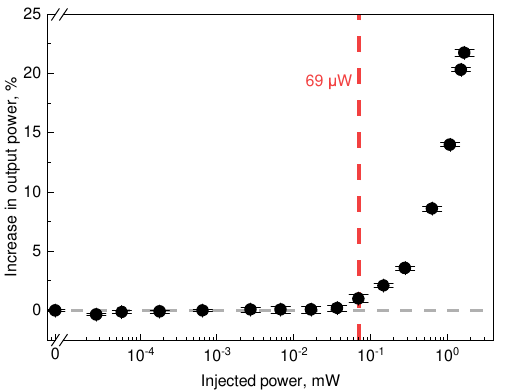}
\caption{Average output power of pulsed LD1550 under exposure to $1310$-$\nano\meter$ light. }\rev{Standard deviation is less than $1\%$ of measured average power.}
\label{fig:avgpower}
\end{figure}

This behaviour might result in an increase of the mean photon number emitted by Alice. To estimate the effect of optical pumping on QKD security, we measure LD1550's output characteristics in the pulsed regime in the presence of attack. \Cref{fig:avgpower} shows the increase in the average output power of Alice's laser when injecting a different amount of power at $1310~\nano\meter$. A notable and stable increase takes place when Eve's pump power reaches $69~\micro\watt$. With a further pump power increase, the LD1550's power rises linearly. It reaches $21.7\%$ at pump power of $1.6~\milli\watt$. 

\begin{figure}
\includegraphics{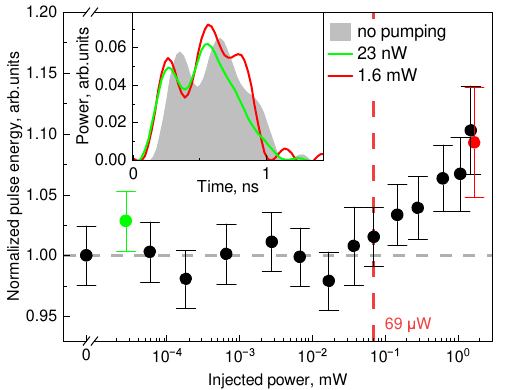}
\caption{Pulse energy of LD1550's pulses under Eve's illumination, normalised to their energy without pumping. \rev{Error bars present the standard deviation.} Inset shows typical single-shot pulse shapes before exposure and under exposure to the minimum and maximum 1310-nm laser powers. The pulses change shape. Their mean timing, however, does not change; the time shift visible in the plot is the result of a random jitter of individual pulses.}
\label{fig:area}
\end{figure}

We also measure the shape of pulses emitted by Alice under Eve's illumination. For each experimental point, we \rev{record 200k pulses, calculate the mean pulse area and its standard deviation, and draw} a normalised pulse energy as the pulse area under exposure divided by the initial pulse area before exposure. The result is shown in \cref{fig:area}. Here, the maximum increase in pulse energy is about $10\%$ at the maximum LD1310 power. A stable increase in the pulse energy is observed starting from $140~\micro\watt$ pump power. However, the pulse shape changes even under the lowest applied pump power of just $23~\nano\watt$ (inset in \cref{fig:area}). %The pulses have jitter of about $75~\pico\second$, which manifests in different time positions of the single-shots.
\rev{The observed shifting of pulses under attack by about $75~\pico\second$ is an order of measurement accuracy.}

The behaviour of the pulse energy differs quantitatively from that of the average output power. In~\cref{fig:area}, even a decrease of the pulse energy is observed. This discrepancy may be caused by spontaneous luminescence of LD1550 under continuous pumping by LD1310, which cannot be distinguished in our measurements. 

In summary, our experimental results show that just about $23~\nano\watt$ of $1310$-$\nano\meter$ light might change characteristics of Alice's pulses, while about $140~\micro\watt$ should reach her laser diode to induce an increase in its pulse energy.

\section{Discussion}
\label{sec:discussion}

As shown in~\cref{sec:results}, optical pumping induces an increase of both the average power and pulse energy. Let us discuss the implications of each for QKD security.

\medskip{\em Increase of average power.} In our experiment done at a low duty cycle, we notice that the increase in average power under pumping is higher than the increase in pulse energy. The difference reaches $10\%$ under the highest attacker's power. This can be explained by an emission of LD1550 between signal pulses, induced by the optical pumping. However, the emission power between the pulses is about 1000 times lower than in the pulses. It might be difficult for Eve to exploit.

\medskip{\em Increase of pulse energy.} This increases the mean photon number emitted. Its effect on the security of different QKD protocols is well-studied, particularly for the standard decoy-state BB84 and MDI QKD protocols \cite{huang2019}. For instance, in a typical BB84 system \cite{comandar2016}, an unaccounted increase of the pulse energy by 10\% leads to an overestimate of the secret key rate by 11\% at the distance of $40~\kilo\meter$ of fiber \cite{huang2019}, which makes the QKD system insecure. \rev{Hereinafter, we present calculations indicating that the attack can be entirely mitigated by employing sufficient attenuation of the eavesdropper's light. With the implemented countermeasure, this attack will not increase Alice's pulse intensity and affect the secret key rate.}%\rev{In this work we do not provide any security evaluate, this will be investigate in next works. For this case we consider that is amount of isolation, calculated in this work, is enough to reduce adjustemnts to secret key ratio are small enough to be neglected.}

\medskip{\em Countermeasures.} To prevent the optical-pumping attack, different known techniques might be considered. \rev{They include \rev{real-time calibration of Alice's intensity using variable optical attenuators with feedback}, monitoring incoming light from the quantum channel, using optical power limiting devices~\cite{zhang2021,peng2023,lovic2023}, and providing a sufficiently high total isolation~\cite{ponosova2022} to suppress the injected light from the quantum channel to a safe level}.

In our experiment, the $1310$-$\nano\meter$ pump power required to observe a stable increase of both differential quantum efficiency and $1550$-$\nano\meter$ pulse energy is about $140~\micro\watt$ ($-8.5~\deci\bel\milli$). It is several orders of magnitude higher comparing to the laser-seeding attacks \cite{huang2019,pang2020,lovic2023}. Therefore, the optical-pumping attack should be easily preventable by a proper isolation level. %\rev{Real-time calibration might not be a perfect solution because of this method changes parameters of laser slowly. We believe that the delay in the reaction of the carrier concentration to optical pumping is of the order of the intraband relaxation time, which is of the order of $0.1\pico\second$~\cite{agrwal1988}. The lifetime of carriers created by optical-pumping is required.}

A major limiting factor for Eve is the power-handling ability of the quantum channel and of the last component in the QKD source setup. Owing to the lack of experimental and theoretical data on this, we assume that the last component in the QKD source is a fiber-optic isolator and its damage threshold at $1310~\nano\meter$ equals that at $1550~\nano\meter$, which is on order of $4~\watt$ ($36~\deci\bel\milli$)~\cite{ponosova2022}. Then, Alice needs isolation at $1310~\nano\meter$ just above $44.5~\deci\bel$ to prevent the optical-pumping attack. Meanwhile, a Raman fiber laser \rev{based on a standard single-mode fiber (OFS SMBD0980B)} of about $250~\watt$ cw power at $1310~\nano\meter$ is reported in~\cite{grimes2022}. In this case, the required isolation is $62.5~\deci\bel$.
%However, in both cases, it is significantly lower than the isolation required in Alice's module in order to guarantee its security against the Trojan-horse attack, which is roughly $170~\deci\bel$ or higher~\cite{lucamarini2015}. Therefore, we might conclude that any QKD system with proper protection against the Trojan-horse attack by isolation components will be resilient to the optical-pumping attack. 
%However, if the QKD system lacks modulators, it is not susceptible to the Trojan-horse attack and might not have enough isolation installed, such as passive-state-preparation schemes \cite{comandar2016,yaun2016,roberts2018,paraiso2019,victor2023,wang2023,kurochkin2024}. Then it's important to ensure it is protected against the optical-pumping attack.

\begin{figure}
\includegraphics{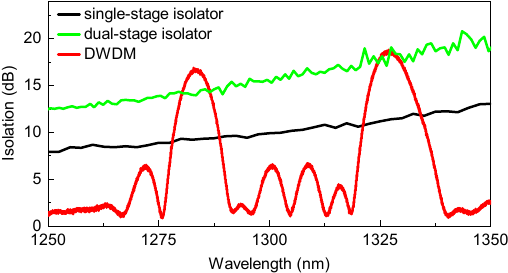}
\caption{Spectral characteristics near $1310~\nano\meter$ of typical $1550$-$\nano\meter$ fiber-optic isolators in backward direction and a dense-wavelength-division multiplexer (DWDM; Prointech DWDM-1T-MOD777-34) measured from its common port to the port of channel 34.}
\label{fig:spectra}
\end{figure}

 \rev{The estimated safe isolation boundaries are significantly lower than the typical isolation at $1550~\nano\meter$ of practical active-state-preparation Alice modules in their backward direction~\cite{tan2021,sajeed2021,makarov2024}. Unfortunately, systems might be vulnerable to the optical-pumping attack owing to the spectral dependency of the isolation of an optical scheme.} 
Some of the passive elements used to prevent attacks, such as fiber-optic isolators and dense-wavelength-division multiplexers (DWDMs), often have vulnerabilities at wavelengths differing from their operating one \cite{jain2015,Nasedkin2022,Borisova2020,tan2025,nasedkin2023}. \Cref{fig:spectra} shows typical wavelength-dependence of loss of $1550$-$\nano\meter$ telecommunication fiber-optic isolators and DWDM near $1310~\nano\meter$. At $1550~\nano\meter$, a typical single-stage isolator provides isolation of about $30~\deci\bel$ \cite{ponosova2022} and dual-stage isolator of about $50~\deci\bel$. At $1310~\nano\meter$, they provide isolation of only about $10	$ and $15~\deci\bel$. DWDM filters also have the same issue with the lack of isolation outside their operating range, as shown in~\cref{fig:spectra}. Their isolation at $1310~\nano\meter$ between the common input and channel output fluctuates around a few decibel. So, the spectral dependence of the system components might open a loophole for Eve to conduct the optical-pumping attack and should be considered in the system design. 

\rev{However, if the QKD system lacks modulators, it is not susceptible to the light-injection attacks and might not have enough isolation installed, such as passive-state-preparation schemes \cite{comandar2016,yaun2016,roberts2018,paraiso2019,zapatero2023,wang2023,kurochkin2024}. Then it's important to ensure it is protected against the optical-pumping attack.}

\section{Risk evaluation for a practical QKD implementation}
\label{sec:real_QKD}

\begin{figure}
\includegraphics{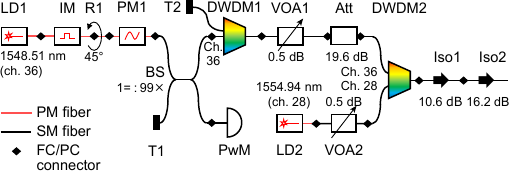}
\caption{Optical scheme of a commercial QKD transmitter~\cite{makarov2024}. \rev{LD, laser diode;} IM, intensity modulator; R, FC/PC connector with $45\degree$ rotation; PM, phase modulator; T, optical terminator; BS, beamsplitter; \rev{DWDM, dense-wavelength-division multiplexer; Ch., DWDM channel number; VOA, variable optical attenuator; PwM, power meter;} Att, fixed attenuator; Iso, polarisation-independent isolator.}
\label{fig:QRate-setup}
\end{figure}

We estimate the success of the optical-pumping attack on an industrial-prototype prepare-and-measure QKD system, on the example of a real optical scheme of Alice produced by QRate \rev{~\cite{QKD312}} that is analysed in detail in~\cite{makarov2024}. \Cref{fig:QRate-setup} shows it. Alice uses intensity and phase modulators IM and PM1 to prepare her states, and isolators Iso1 and Iso2 to protect them against the Trojan-horse attack. Laser diode LD1 emits signal pulses and is the target of our attack. The following calculations consider the optical path of Eve's light to it. The total isolation at $1310~\nano\meter$ is calculated as a sum of loss in backward direction of each component, and is
\begin{equation}
\begin{aligned}
\alpha_{1310} = ~&\alpha_\text{Iso2} + \alpha_\text{Iso1} + \alpha_\text{DWDM2} + \alpha_\text{Att} + \alpha_\text{VOA1}\\
	&+ \alpha_\text{DWDM1} + \alpha_\text{BS} + \alpha_\text{PM1} + \alpha_\text{IM} + \alpha_\text{LD1},
\end{aligned}
\label{eq:power-pump}
\end{equation}
where $\alpha_\text{Iso}$ is the isolation value of the optical isolator at $1310~\nano\meter$ wavelength, $\alpha_\text{LD1}$ is the isolation of LD1's built-in isolator, $\alpha_\text{DWMD}$, $\alpha_\text{Att}$, $\alpha_\text{VOA1}$, $\alpha_\text{BS}$, $\alpha_\text{PM1}$, and $\alpha_\text{IM}$ are insertion losses of components in \cref{fig:QRate-setup}.

\begin{table}[b]
\vspace{-0.7em} % compensates for REVTeX layout bug
\caption{Insertion loss of components similar to those from the QKD system~\cite{makarov2024} measured at $1310~\nano\meter$. The variable optical attenuator can be set
anywhere in the range $0.5$--$30~\deci\bel$, of which the worst case of $0.5~\deci\bel$ is assumed here.}
\begin{tblr}[t]{llr}
\hline\hline
Element &Symbol &Loss, dB \\
\hline
Isolator 2 &$\alpha_\text{Iso2}$ &16.2 \\
Isolator 1 &$\alpha_\text{Iso1}$ &10.6 \\
DWDM2 &$\alpha_\text{DWMD2}$ &3.0 \\
Fixed attenuator &$\alpha_\text{Att}$ &19.6 \\
Variable optical attenuator &$\alpha_\text{VOA1}$ &0.5 \\
DWDM1 &$\alpha_\text{DWMD1}$ &4.1 \\
Beamsplitter &$\alpha_\text{BS}$ &24.0 \\
Phase modulator &$\alpha_\text{PM1}$ &4.5 \\
Intensity modulator &$\alpha_\text{IM}$ &4.5 \\
LD1's built-in isolator &$\alpha_\text{LD1}$ &10.6 \\
\hline\hline
\end{tblr}
\label{tab:isolation1310}
\end{table}

The isolation values at $1310~\nano\meter$ used for the calculation in~\cref{eq:power-pump} are listed in~\cref{tab:isolation1310}. To estimate Alice's setup isolation at $1310~\nano\meter$, we measure insertion loss of components similar to those listed in~\cite{makarov2024}. We cannot disclose model numbers for most of them, owing to our confidentiality agreements with QKD system manufacturers. They are standard off-the-shelf fiber-optic products. Fixed attenuator (Thorlabs FA20T) and fiber-optic 99:1 beamsplitter (Thorlabs TW1550R1A2) specify loss at $1310~\nano\meter$ in their data sheets. A phase modulator based on Ti-diffused lithium niobate is characterised using our PwM and LD1310. Insertion loss of IM is assumed to be the same as that of PM. All the other components are characterised using a broadband light source and optical spectrum analyser (Hewlett-Packard 70004A), using methodology from Appendix~E of~\cite{makarov2024}. The isolation of LD1's built-in isolator is assumed to be the same as that of the single-stage isolator.

The total calculated isolation at $1310~\nano\meter$ is $97.6~\deci\bel$, which is higher comparing to our maximum required value of $62.5~\deci\bel$. Thus, the system is resilient against the optical-pumping attack at $1310~\nano\meter$. However, to ensure its security, both the efficiency of optical pumping and insertion loss of the source components must be characterised in a very wide spectral range~\cite{tan2025,makarov2024}.

\section{Conclusion}
\label{sec:conclusion}

We have proposed a new kind of attack on real QKD systems---the optical-pumping attack on the transmitter. It allows the eavesdropper to increase the intensity of the prepared states by injecting light into Alice at a wavelength corresponding to a semiconductor absorption band of her laser source. We experimentally demonstrate 10\% increase in pulse energy of $1550$-$\nano\meter$ Alice's source using Eve's injected power of $1.6~\milli\watt$ at $1310~\nano\meter$. 

Our study shows that the power required for the success of this attack is at least three orders of magnitude higher than that of the laser-seeding attack. At the same time, characteristics of passive countermeasures in practical QKD systems are wavelength-dependent and might be ineffective against this type of attack in a wide spectral range. Thus, the optical-pumping attack should be considered a possible threat to QKD security. \rev{As part of the certification process, QKD systems must be tested to confirm their countermeasures effectively mitigate this attack.}%\rev{For certification purposes of any system it is necessary to calculate the isolation level that is required to prevent this attack.} 

Finally, we analyse the risk of this attack on the example of the industrial QKD system\rev{~\cite{QKD312,makarov2024}}. The analysis indicates that systems with proper protection against the \rev{light-injection attacks} may be resilient against the optical-pumping attack \rev{with existing countermeasures}.%The analysis indicates that systems with proper protection against the Trojan-horse attack may also be expected to be resilient against the optical-pumping attack.
 Therefore, the latter should be strongly considered in QKD systems that do not require protection against the \rev{light-injection attacks}, such as the systems using passive state preparation \cite{comandar2016,yaun2016,roberts2018,paraiso2019,zapatero2023,wang2023,kurochkin2024,hu2023, lu2023}.

%\acknowledgments

\bigskip
{\em Funding:} Russian Science Foundation (grant 21-42-00040). Q.P.\ and A.H.\ acknowledge funding from the National Natural Science Foundation of China (grant 62371459) and the Innovation Program for Quantum Science and Technology (grant 2021ZD0300704), and Postdoctoral Fellowship Program of CPSF under Grant Number GZC20252817. V.M.\ acknowledges funding from the Galician Regional Government (consolidation of research units: atlanTTic and own funding through the ``Planes Complementarios de I+D+I con las Comunidades Aut{\' o}nomas'' in Quantum Communication), MICIN with funding from the European Union NextGenerationEU (PRTR-C17.I1), and the ``Hub Nacional de Excelencia en Comunicaciones Cu{\' a}nticas'' funded by the Spanish Ministry for Digital Transformation and the Public Service and the European Union NextGenerationEU.

\medskip
{\em Author contributions:} \rev{R.S.,\ A.H.,\ and Q.P.\ assisted in planning the experiment.} M.F.\ and A.P.\ conducted the experiment. R.S.\ and V.M.\ supervised the study. All authors analysed the results and contributed to writing the manuscript.

\medskip

{\em Disclosures:} The authors declare no conflicts of interest.

\medskip

{\em Data availability:} Data underlying the results presented in this paper are not publicly available at this time, but may be obtained from the authors upon reasonable request.
%\begin{backmatter}
%\bmsection{Funding} Content in the funding section will be generated entirely from details submitted to Prism. Authors may add placeholder text in the manuscript to assess length, but any text added to this section in the manuscript will be replaced during production and will display official funder names along with any grant numbers provided. If additional details about a funder are required, they may be added to the Acknowledgments, even if this duplicates information in the funding section. See the example below in Acknowledgements. For preprint submissions, please include funder names and grant numbers in the manuscript.

%\bmsection{Acknowledgments} Additional information crediting individuals who contributed to the work being reported, clarifying who received funding from a particular source, or other information that does not fit the criteria for the funding block may also be included; for example, ``K. Flockhart thanks the National Science Foundation for help identifying collaborators for this work.''

%\bmsection{Funding} ...

%\smallskip

%\bmsection{Disclosures} The authors declare no conflicts of interest.

%\smallskip

%\bmsection{Data availability} Data underlying the results presented in this paper are not publicly available at this time, but may be obtained from the authors upon reasonable request.

%\end{backmatter}

%\def\bibsection{\medskip\begin{center}\rule{0.5\columnwidth}{.8pt}\end{center}\medskip} % Redefines bibliography separator to single-column. This reduces chances of layout bugs in the last page.

\appendix
\setcounter{figure}{0}
\setcounter{section}{0}
\setcounter{table}{0}
\numberwithin{figure}{section}
\numberwithin{table}{section}
\numberwithin{equation}{section}

\section{Supplementary materials}
\label{sec:supplementary}

Distributed feedback (DFB) laser diodes are widely used in quantum key distribution (QKD) systems because they offer low noise, high-frequency stability, and a narrow linewidth. The DFB laser achieves high-performance emission by combining an active medium (gain semiconductor) with a diffraction grating along the entire length of the cavity. This design allows for precise selection of the wavelength.

\begin{figure}[ht!]
\includegraphics{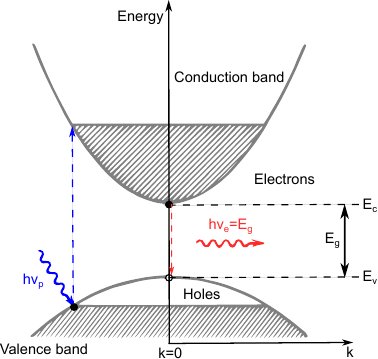}
\caption{Schematic view of the conduction and valence bands in semiconductor material. The pump light is in blue, and the emission light is in red. $E_c$ represents the energy level at the bottom of the conduction band, and $E_v$ the energy level at the top of the valence band.}
\label{fig:bands}
\end{figure}

An injection current transfers carriers to the active region of LD. When this current reaches a sufficient level to create a population inversion, lasing—or stimulated emission—occurs as a result of carrier recombination: electrons from the conduction band recombine with holes in the valence band. Optical pumping excites electrons from the valence band into the conduction band inside the active region of LD. \Cref{fig:bands} provides a schematic illustration of the conduction and valence bands. For a detailed explanation of the actual band structure in InGaAsP materials typically applied in 1550-nm laser diodes, refer to~\cite{goldberg1999}.

In the presence of injection current alone, the laser rate equation for the carrier number $N$ is expressed as follows:
\begin{equation}
	\begin{split}
		\dot{N}&=I/e-N/\tau_e-QG/(\Gamma\tau_\text{ph}).
	\end{split}
\end{equation}
The effect of continuous optical pumping at a wavelength of 1310\,nm is reduced to the addition of the optical pumping rate $R_\text{opt}$ to the right hand side of the laser rate equation for the carrier number $N$. The system of rate equations can thus be written as follows:
\begin{equation}
	\begin{split}
		\dot{N}&=I/e+R_\text{opt}-N/\tau_e-QG/(\Gamma\tau_\text{ph}),\\
		\dot{Q}&=(G-1)Q/\tau_\text{ph}+C_\text{sp}N/\tau_e,
	\end{split}
\end{equation}
where $Q$ is the normalized electric field intensity corresponding to the photon number inside the laser cavity and related to the output power by ${P=Q\eta\hbar\omega_0/(2\Gamma\tau_\text{ph})}$, where $\hbar\omega$ is the photon energy ($\omega$ is the central angular frequency of the 1550-nm laser), $\eta$ is the differential quantum output, $\Gamma$ is the confinement factor, $\tau_\text{ph}$ is the photon lifetime inside the cavity, and the factor $1/2$ takes into account that the output power is measured only from one facet. Onwards, $I$ is the pump current, $e$ is the absolute value of the electron charge, $\tau_e$ is the effective lifetime of the electron, the factor $C_\text{sp}$ corresponds to the fraction of spontaneously emitted photons that end up in the active mode, and the dimensionless gain $G$ is defined by
\begin{equation}
	G=\frac{N-N_0}{N_\text{th}-N_0}\frac{1}{\sqrt{1+2\gamma_Q Q}},
\end{equation}
where $N_0$ and $N_\text{th}$ are the carrier numbers at transparency and threshold, respectively, and $\gamma_Q$ is the dimensionless gain compression factor. The optical pumping rate, in turn, can be written as
\begin{equation}
	R_\text{opt}=\epsilon_\text{opt}\frac{P_{1310}}{\hbar\omega_{1310}},
\end{equation}
where $\epsilon_\text{opt}$ is the pumping efficiency,  $P_{1310}$ is the optical pumping power, and $\hbar\omega_{1310}$ is the corresponding photon energy.

\begin{figure}
\includegraphics{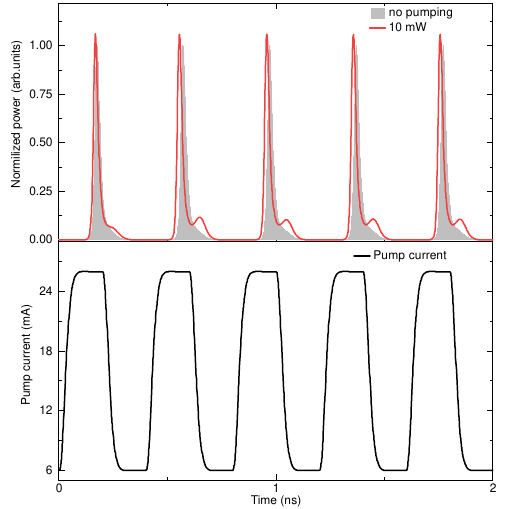}
\caption{Simulations of the output signal with and without optical pumping.}
\label{fig:sim}
\end{figure}

The actual characteristics of laser diodes required for near-practical simulation, such as the absorption cross-sections of the semiconductor, the material properties, and the coupling efficiency with optical fibre, are unknown because manufacturers keep this information confidential. For simulations we have used laser and pump current parameters listed in~\cref{tab:SimParameters}.
Simulations of the output signal with and without optical pumping are shown in~\cref{fig:sim}. It was assumed that the pump current is a sequence of rectangular pulses and can be written as $I(t)=I_\text{b}+I_\text{p}(t)$, where $I_\text{b}$ is the bias current, and the modulation current $I_\text{p}(t)$ varied from 0 to $I_p^\text{max}$ (the peak-to-peak value of the modulation current). 

\begin{table}
	\caption{Simulation parameters.}
	\label{tab:SimParameters}
	\begin{center}		
		\tabcolsep=3pt		
		\begin{tabular}{lc}
			\hline
			\hline
			\rule{0mm}{3mm}
			Parameter & Value\\
			\hline
			\rule{0mm}{3mm}
			Bias current $I_\text{b}$,\,mA &6.0\\
			\rule{0mm}{2mm}
			Maximum pump current $I_\text{p}^\text{max}$,\,mA &20.0\\
			\rule{0mm}{3mm}
			Carrier timelife $\tau_e$, ns &1.0\\
			\rule{0mm}{3mm}
			Photon timelife $\tau_\text{ph}$, ps &3.0\\
			\rule{0mm}{3mm}
			Pump current pulse width, ns &0.2\\
			\rule{0mm}{3mm}
			Pulse repetition rate, GHz &2.5 \\
			\rule{0mm}{3mm}
			Confinement factor $\Gamma$ &0.12\\
			\rule{0mm}{3mm}
			Threshold carrier number $N_\text{th}$ &$6.5 \times 10^7$\\
			\rule{0mm}{3mm}
			Transparency level $N_{0}$ &$5.5\times10^7$\\
			\rule{0mm}{3mm}
			Spontaneous emissions fraction $C_\text{sp}$ &$10^{-5}$\\
			\rule{0mm}{3mm}
			Gain compression factor $\gamma_Q$ &1.0$\times10^{-6}$\\
			\rule{0mm}{3mm}
			Pumping efficiency $\epsilon_\text{opt}$ &0.1 \\
			\hline
			\hline
		\end{tabular}
	\end{center}	
\end{table}

% Bibliography

% Full bibliography added automatically for Optics Letters submissions; the following line will simply be ignored if submitting to other journals.
% Note that this extra page will not count against page length
%\bibliographyfullrefs{library}

\def\bibsection{\medskip\begin{center}\rule{0.5\columnwidth}{.8pt}\end{center}\medskip} % Redefines bibliography separator to single-column. This reduces chances of float placement bugs in the last page.
\bibliography{library}

\end{document}